\documentclass{iopart}

\begin{document}

\newcommand{\cfield}{\mathbb{C}}
\newcommand{\rfield}{\mathbb{R}}
\newcommand{\ad}{\mbox{\rm ad}}
\newcommand{\su}{\mbox{\rm su}}
\newcommand{\sop}{\mbox{\rm sp}}
\newcommand{\du}{\mbox{\rm u}}
\newcommand{\SU}{\mbox{\rm SU}}
\newcommand{\U}{\mbox{\rm U}}
\newcommand{\GL}{\mbox{\rm GL}}
\newcommand{\gl}{\mbox{\rm gl}}
\newcommand{\dm}{\mbox{\rm dim}}
\newcommand{\eff}{\mbox{\scriptsize \rm  eff }}
\newcommand{\total}{\mbox{\scriptsize \rm  total }}
\newcommand{\gen}{\mbox{ \rm  gen }}
\newcommand{\ddet}{\mbox{ \rm  det }}
\newcommand{\lie}{{\cal L}}
\newcommand{\im}{{\mathrm i}}

\title{Indirect controllability of degenerate quantum systems with quantum accessor}
\author{Si Li, Z. F. Jiang and H. C. Fu \footnote{
E-mail: hcfu@szu.edu.cn, corresponding author.}}

\address{School of Physical Sciences and Technology, Shenzhen University, \\
Shenzhen 518060, P.\,R.\,China}

\begin{abstract}
Complete controllability of degenerate quantum system using quantum accessor
modeled as a qubit chain with nearest neighborhood coupling is investigated.
Sufficient conditions on the length of accessor and the way of
coupling between controlled system
and accessor are obtained. General approach to arbitrary finite system is
presented and two and three level degenerate systems are investigated
in detail.
\end{abstract}
\maketitle

\section{Introduction}

Quantum control is a coherent or an incoherent process to
steer a quantum system to a given target state \cite{1}.
It was first proposed by Huang \cite{1} and Belavkin \cite{2} {\it et.~al.}
and is significant in many fields of physics and chemistry, especially
in the quantum computation and quantum information
processing \cite{ref1}. Actually the universality of quantum logic gates can be
understood from viewpoint of complete controllability in quantum control \cite{universality}. Conventional
quantum control is the coherent control of quantum systems using classical external fields.
Controllability of this semi-classical control is well studied \cite{controllability}, especially the complete
controllability of finite-dimensional quantum systems using Lie algebra approach \cite{fu, schirmer},
graph \cite{graph1} and transfer graph approach \cite{graph2}. The Lie algebra approach plays an
important role in the investigation of both the classical control \cite{lie-approach} and the quantum control.

In some circumstances in quantum computation, there is need to control qubits
using quantum controllers such as a quantum accessor or its environment, for example,
in the study of fundamental limit of quantum information processing and influence
of decoherence to quantum control \cite{xue}, the so-called encoded qubits to realize the universal quantum
computation with local manipulation of physical qubits \cite{zhou},
the universal indirect control of nuclear spins using a single electron spin
acting as an accessor driven by microwave irradiation of resolved anisotropic hyperfine \cite{hodges},
incoherent control induced by
environment modeled as quantum radiation fields \cite{romano,romano2},
and so on.
Therefore the control of quantum systems using quantum controllers has significant application
in quantum information processing and has attracted much attention recently. Recently,
D'Alessandro {\it et.~al.} investigated the exact algebraic conditions for indirect controllability
of finite quantum system \cite{indirect-condition}.

Authors of
this paper proposed a scheme for the control of an arbitrary finite
degenerate quantum systems using a quantum accessor modeled as a qubit chain with
$XY$-type neighborhood coupling and the clasical control fields control
the accessor only \cite{fu-indirect1,fu-indirect2}.
On the other hand, we investigated the complete controllability of quantum
system with two-fold degeneracy \cite{zd}. In this paper, based on
those works, we shall investigate
the indirect controllability of degenerate quantum systems through quantum accessor.
We find the conditions for the complete controllability and it depends on the way
of coupling between the system and accessor and the length of qubit chain.

In this paper we use $\im \equiv \sqrt{-1}$.

\section{Control system}

Consider a finite-dimensional quantum system $S$ with $N$ energy
levels, described by the Hamiltonian
\begin{eqnarray}
H_S =  \sum_{n=1}^{N}\sum_{i=1}^{\beta_n} E_n e_{ni,ni},
\end{eqnarray}
where $e_{ij,kl}\equiv |i,j\rangle\langle k,l|$ and $|i,j\rangle$ is
the $j$-th degenerate state of the $i$-th level,
$\beta_n$ is the degree of degeneracy of the level $n$.
Suppose that the ground state is non-degenerate, namely $\beta_1=1$,
and $\tr H_S=0$, without losing generality. The Hilbert space of
the controlled system is of dimension
\begin{equation}
{\cal N} =   \sum_{i=1}^{N}\beta_i.
\end{equation}

We would like to control this quantum system by an quantum accessor.
As in previous papers \cite{fu-indirect1,fu-indirect2}, we model the accessor as a qubit chain
with $XY$-type coupling
\begin{eqnarray}
 H_{A} = H_A^0 + H_A^I, \quad H_A^0 = \sum_{k=1}^{M} \hbar \omega_{k} \sigma_{z}^{k},  \quad
 H_A^I = \sum_{k=1}^{M-1}c_{k} \sigma_{x}^{k}\sigma_{x}^{k+1} ,
\end{eqnarray}
where $H_{A}^{0}$ is the free Hamiltonian of accessor,
$H_{A}^{I}$ is interaction Hamiltonian with nearest neighborhood coupling,
coupling constant $c_{j}\neq 0$, and $\sigma_{\alpha}^{j}\ (\alpha=x,y,z)$
is the Pauli matrix of the $j$-th qubit
\begin{eqnarray}\label{eqR:Pauli}
\sigma_{\alpha}^{j}=\underbrace{1  \otimes \cdots \otimes 1}_{j-1} \otimes
\sigma_{\alpha}^{j} \otimes \underbrace{1 \otimes \cdots \otimes 1}_{M-j},
\qquad \alpha=x,y,z.
\end{eqnarray}

Interaction Hamiltonian between the system $S$ and accessor is generally written as
\begin{eqnarray}
H_{SA} =  \sum_{\{ \alpha_{p}\}} \left[ \sum_{n=1}^{N-1}
\sum_{i=1}^{\beta_n}\sum_{j=1}^{\beta_{n+1}}\sum_{k=0,\pm1}
g_{\{\alpha_{p}\}}^{nij(k)} s_{nij}^{k} \right.
\otimes \left. \left( \prod_{p=1}^{M} \sigma_{\alpha_{p}}^{p} \right)\right],
\end{eqnarray}
where $g_{\{\alpha_{p}\}}^{nij(k)}$ are coupling constants,
$\{ \alpha_{p}\}=\{ \alpha_{1},\alpha_{2},\cdots,\alpha_{M}\}$,
and each $\alpha_{p}=x, y, z$.
System operators $s_{nij}^{k}$ are defined as
\begin{eqnarray}\label{eqR:id3Sk}
s^{k}_{nij}= \left\{
\begin{array}{ll}
x_{ni,(n+1)j}, & k=1; \\
h_{ni,(n+1)j}, & k=0; \\
y_{ni,(n+1)j}, & k=-1,
\end{array} \right.
\end{eqnarray}
where
\begin{eqnarray}
& x_{ni,(n+1) j}=e_{ni,(n+1) j} + e_{(n+1) j, ni},\nonumber \\
& h_{ni,(n+1)j}=e_{ni,ni} - e_{(n+1)j,(n+1)j},\nonumber \\
& y_{ni,(n+1)j}=i(e_{ni,(n+1)j} - e_{(n+1)j,ni}),
\end{eqnarray}
are Hermitian operators and $e_{ni,mj}=|{ni}\rangle\langle mj|$.
We note that skew-hermitian
operators $\im x_{ni,(n+1) j}, \im y_{ni,(n+1) j}$ and $\im h_{ni,(n+1) j}$ form
the Chevalley basis of $\su ({\cal N})$ Lie algebra.

We use two independent classical fields $f_{i}(t)$ and
$f_{i}^{\prime}(t)$, $i=1,2,\cdots,M$, to control each qubit. Then the total
Hamiltonian of the control system reads
\begin{eqnarray}
& H = H_{0} + \sum_{i=1}^{M}\left[ f_{i}(t) \sigma_{x}^{i} +
  f_{i}^{\prime}(t) \sigma_{y}^{i}\right], \\
& H_{0} = H_S + H_A + H_{SA}.
\end{eqnarray}

To examine the complete controllability, we need
to prove that the Lie algebra $\lie$ generated by
\begin{eqnarray}
\lie = \mbox{\rm gen}\left\{ \left.
\im H_0, \im 1\otimes\sigma_{x}^{k}, \im 1\otimes\sigma_{y}^{k} \
\right| \  1\leq k \leq M \right \}
\end{eqnarray}
is $\su({\cal N} 2^M)$.

It is easy to see that there are $ \tilde{N}\equiv 3 \sum_{n=1}^{N-1}\beta_n \beta_{n+1}$ elements $s_{nij}^k$
and $3^M$ elements $\left( \prod_{p=1}^{M} \sigma_{x}^{p} \right)$
in the interaction hamiltonian $H_{SA}$.
To investigate the complete
controllability, we would like to split the interaction Hamiltonian $\im H_{SA}$
to obtain each $ \im s_{nij}^{k} \otimes \left( \prod_{p=1}^{M} \sigma_{x}^{p} \right)\in \lie$.
As discussed in \cite{fu-indirect2}, the number $M$ of qubits in accessor has to satisfy
the condition
\begin{equation}
3^M \geq \tilde{N}. \label{chain-length-condition}
\end{equation}
This requirement is true for both non-degenerate and degenerate systems
as it is only related to the numbers of Lie algebra generators.

In the following we shall first discuss the two- and three-dimensional
systems as explicit examples in Sec.~3 and 4 respectively,
and then present a general approach to arbitrary finite dimensional systems in Sec.~5.

\section{Controllability of a two-level system}

For two level system with non-degenerate ground state and
two-bold degenerate excited states, we have proved that it
is not completely controllable in the semi-classical control scheme
\cite{zd}. We naturally ask
whether it is possible to control it using the indirect control scheme. We shall answer
this question in this section.

Consider a two-level system with the non-degenerate ground state
and two-fold degenerate excited states described by the Hamiltonian
\begin{eqnarray}\label{eqR:id2Hs}
H_S = E_1 e_{11,11} + E_2 e_{21,21} + E_2 e_{22,22}
    = \varepsilon_{1} h_{11,21} + \varepsilon_{1} h_{11,22},
\end{eqnarray}
where $E_1$, $E_2$ are eigen-energy, $e_{ni,ni}=| n,i \rangle \langle n,i |$ are
eigen-states and $i$ denotes different degenerate state, and
$\varepsilon_{1} = 2^{-1}E_1$. Here we suppose $\tr H_S=E_1 + 2 E_2=0$ without
losing generality.
From condition (\ref{chain-length-condition}),
the accessor consists of two qubits and its Hamiltonian is
\begin{eqnarray}
H_A = \hbar \omega_{1} \sigma_{z}^{1} + \hbar \omega_{2}
\sigma_{z}^{2} + c \sigma_{x}^{1} \sigma_{x}^{2}.
\end{eqnarray}
Interaction Hamiltonian between the system and accessor is explicitly written as
\begin{eqnarray}
H_{SA}
=& \sum_{ \alpha,\beta=x,y,z} \left(  g_{\alpha\beta}^{1(0)} h_{11,21}
+ g_{\alpha\beta}^{2(0)} h_{11,22}+ g_{\alpha\beta}^{1(1)} x_{11,21}
 \right. \nonumber \\
&  \left.
   + g_{\alpha\beta}^{2(1)} x_{11,22}
   + g_{\alpha\beta}^{1(-1)} y_{11,21}
   + g_{\alpha\beta}^{2(-1)} y_{11,22} \right)
\otimes  \sigma_{\alpha}^{1}\sigma_{\beta}^{2}.
\end{eqnarray}

The Hamiltonian of the total control system is
\begin{eqnarray}
H = H_S  + H_A + H_{SA} + \sum_{i=1}^{2}
\left[ f_{i}(t)\sigma_{x}^{i} + f_{i}^{\prime}(t) \sigma_{y}^{i}\right].
\end{eqnarray}

Let $\lie$ be the Lie algebra generated by
$\{ \im H_0, \im 1 \otimes \sigma_{x}^{i}, \im 1\otimes\sigma_{y}^{i}\ |\ i=1,2\}$.
It is easy to see that
\begin{equation}
 -\frac{1}{2}\left[\im 1 \otimes \sigma_{x}^{i}, \im 1 \otimes \sigma_{y}^{i}\right]=
\im 1 \otimes \sigma_{z}^{i} \in \lie,
\end{equation}
where $i=1,2$.
Subtracting those elements from $H_0$, we obtain
\begin{eqnarray}\label{eqR:idH02}
\im H_0^{\prime} &\equiv&
    \im H_0 - \im \left(\hbar \omega_{1} 1\otimes\sigma_{z}^{1} +
    \hbar \omega_{2} 1\otimes \sigma_{z}^{2}\right) \nonumber \\
&=& \im H_S + \im c \otimes \sigma_x^1 \sigma_x^2 + \im H_{SA} \in \lie.
\end{eqnarray}

Define the selection operators as in \cite{fu-indirect2}
\begin{eqnarray}\label{eqR:idSxy}
& S_{xy}^{k}(*) = \frac{1}{4}[i\sigma_{x}^{k},[i\sigma_{y}^{k},*]], \\
& S_{yx}^{k}(*) = \frac{1}{4}[i\sigma_{y}^{k},[i\sigma_{x}^{k},*]], \\
& S_{zx}^{k}(*)=\frac{1}{4}[i\sigma_{z}^{k},[i\sigma_{x}^{k},*]],\\
& S_{xx}^{k}(*)=-\frac{1}{8}\left[i\sigma_{z}^{k},\left[i\sigma_{x}^{k},[i\sigma_{y}^{k},*]\right]\right],
\end{eqnarray}
with the properties
\begin{eqnarray}
& S_{xy}^{k}(i\sigma_{\alpha}^{k})=
\left\{
\begin{array}{ll}
i \sigma_{y}^{k} & \alpha=x;\\
0 & \alpha=y,z,
\end{array}
\right.
\\
& S_{yx}^{k}(i\sigma_{\alpha}^{k})=
\left\{
\begin{array}{ll}
i \sigma_{x}^{k} & \alpha=y;\\
0 & \alpha=x,z,
\end{array}
\right. \\
& S_{zx}^{k}(i\sigma_{\alpha}^{k})=
\left\{
\begin{array}{ll}
i \sigma_{x}^{k} & \alpha=z;\\
0 & \alpha=y,x.
\end{array}
\right. \\
& S_{xx}^{k}(i\sigma_{\alpha}^{k})=
\left\{
\begin{array}{ll}
i \sigma_{x}^{k} & \alpha=x;\\
0 & \alpha=y,z.
\end{array}
\right.
\end{eqnarray}
Applying selection operators on element (\ref{eqR:idH02}), we have
\begin{eqnarray}
& S_{xx}^{2}S_{yx}^{1}(\im H_0^{\prime}) = \im \left(  g_{yx}^{1(0)} h_{11,21}
+ g_{yx}^{2(0)} h_{11,22}+ g_{yx}^{1(1)} x_{11,21} + g_{yx}^{2(1)} x_{11,22}
\right. \nonumber \\
& \qquad
+ g_{yx}^{1(-1)} y_{11,21}
 \left. + g_{yx}^{2(-1)} y_{11,22} \right)
\otimes  \sigma_{x}^{1}\sigma_{x}^{2} \in \lie,\label{Sxy} \\
& S_{yx}^{2}S_{yx}^{1}(\im H_0^{\prime}) = \im \left(  g_{yy}^{1(0)} h_{11,21}
+ g_{yy}^{2(0)} h_{11,22}+ g_{yy}^{1(1)} x_{11,21}+ g_{yy}^{2(1)} x_{11,22}
 \right. \nonumber \\
& \qquad  \left.
+ g_{yy}^{1(-1)} y_{11,21}
+ g_{yy}^{2(-1)} y_{11,22} \right)\otimes
\sigma_{x}^{1}\sigma_{x}^{2} \in \lie,\label{Syy} \\
& S_{zx}^{2}S_{yx}^{1}(\im H_0^{\prime}) = \im \left(  g_{yz}^{1(0)} h_{11,21}
+ g_{yz}^{2(0)} h_{11,22}+ g_{yz}^{1(1)} x_{11,21}+ g_{yz}^{2(1)} x_{11,22}
 \right. \nonumber \\
&  \qquad \left.
+ g_{yz}^{1(-1)} y_{11,21} + g_{yz}^{2(-1)} y_{11,22} \right)
\otimes  \sigma_{x}^{1}\sigma_{x}^{2} \in \lie.\label{Szy}
\end{eqnarray}
Using $S_{xx}^{2}S_{zx}^{1}$, $S_{yx}^{2}S_{zx}^{1}$, $S_{zx}^{2}S_{zx}^{1}$,
we also have three elements
\begin{eqnarray}\label{S*z}
&\im \left(  g_{z\beta}^{1(0)} h_{11,21}
+ g_{z\beta}^{2(0)} h_{11,22}
 + g_{z\beta}^{1(1)} x_{11,21} + g_{z\beta}^{2(1)} x_{11,22}\right. \nonumber \\
& \qquad \left.
+ g_{z\beta}^{1(-1)} y_{11,21}
 + g_{z\beta}^{2(-1)} y_{11,22} \right)  \otimes  \sigma_{x}^{1}\sigma_{x}^{2} \in \lie, \quad \beta=x,y,z.
\end{eqnarray}
Similarly, we can use selection operators to obtain
\begin{eqnarray}\label{eqR:SL}
& \im \sum_{j=1}^{2}\sum_{k=0,\pm 1}
\left(  g_{x\beta}^{j(k)} s_{j}^{k} \right)
\otimes  \sigma_{x}^{1}\sigma_{x}^{2} \in \lie,\quad \beta=y,z, \label{the9th-element1} \\
& c\im 1\otimes\sigma_{x}^{1}\sigma_{x}^{2} + i\sum_{j=1}^{2}\sum_{k=0,\pm 1}
\left(  g_{xx}^{j(k)} s_{j}^{k} \right)
\otimes  \sigma_{x}^{1}\sigma_{x}^{2} \in \lie, \label{the9th-element}
\end{eqnarray}
where the element (\ref{the9th-element}) has an additional term $c\im 1\otimes\sigma_{x}^{1}\sigma_{x}^{2}$,
which cannot be annihilated by selection operators.

So far we have obtained 9 elements $(\ref{Sxy})$ to $(\ref{the9th-element})$ of $\lie$.
Eight of them from $(\ref{Sxy})$ to $(\ref{the9th-element1})$ are linear combination
of six elements $\{\im h_{11,2i}, \im x_{11,2i}, \im y_{11,2i} | i=1,2\}$ with the same accessor operator $\sigma_x^1 \sigma_x^2$.
Choose any six elements from those nine elements, for example, the first six elements
(from (\ref{Sxy}) to (\ref{S*z})). If their
coefficient determinant is not vanishing,
namely
\begin{eqnarray}\label{eqR:id2det}
\det
\left( \begin{array}{cccccc}
g_{yx}^{1(0)} & g_{yx}^{2(0)} & g_{yx}^{1(1)} & g_{yx}^{2(1)} & g_{yx}^{1(-1)} & g_{yx}^{2(-1)} \\
g_{yy}^{1(0)} & g_{yy}^{2(0)} & g_{yy}^{1(1)} & g_{yy}^{2(1)} & g_{yy}^{1(-1)} & g_{yy}^{2(-1)} \\
g_{yz}^{1(0)} & g_{yz}^{2(0)} & g_{yz}^{1(1)} & g_{yz}^{2(1)} & g_{yz}^{1(-1)} & g_{yz}^{2(-1)} \\
g_{zx}^{1(0)} & g_{zx}^{2(0)} & g_{zx}^{1(1)} & g_{zx}^{2(1)} & g_{zx}^{1(-1)} & g_{zx}^{2(-1)} \\
g_{zy}^{1(0)} & g_{zy}^{2(0)} & g_{zy}^{1(1)} & g_{zy}^{2(1)} & g_{zy}^{1(-1)} & g_{zy}^{2(-1)} \\
g_{zz}^{1(0)} & g_{zz}^{2(0)} & g_{zz}^{1(1)} & g_{zz}^{2(1)} & g_{zz}^{1(-1)} & g_{zz}^{2(-1)}
\end{array}
\right) \neq 0,
\end{eqnarray}
we obtain
\begin{eqnarray}
& \im h_{11,2i} \otimes  \sigma_{x}^{1}\sigma_{x}^{2} \in \lie,\quad
  \im x_{11,2i} \otimes  \sigma_{x}^{1}\sigma_{x}^{2} \in \lie,\nonumber \\
& \im y_{11,2i} \otimes  \sigma_{x}^{1}\sigma_{x}^{2} \in \lie,\quad i=1,2.
\end{eqnarray}
From their commutation relations with $ \im 1 \otimes \sigma_{\alpha}^{k}\in \lie$
($\alpha=x,y,z$),
we obtain elements
\begin{eqnarray}
&  \im h_{11,2i} \otimes  \sigma_{\alpha}^{1}\sigma_{\beta}^{2} \in \lie,\\
&  \im x_{11,2i} \otimes  \sigma_{\alpha}^{1}\sigma_{\beta}^{2} \in \lie,\nonumber \\
&  \im y_{11,2i} \otimes  \sigma_{\alpha}^{1}\sigma_{\beta}^{2} \in \lie,
\ \ i=1,2; \ \alpha,\beta=x,y,z.
\end{eqnarray}
Removing those elements from $\im H_0^\prime$, we obtain
\begin{eqnarray}
&& c^{-1}
\left[
\im H_0^{\prime}  - \im \sum_{ \alpha,\beta=x,y,z} \left(
\sum_{j=1}^{2}\sum_{k=0,\pm 1}
g_{\alpha\beta}^{j(k)} s_{j}^{k} \right)
\otimes  \sigma_{\alpha}^{1}\sigma_{\beta}^{2} \right] \nonumber \\
&& \qquad = \im 1 \otimes \sigma_{x}^{1}\sigma_{x}^{2}\in \lie,
\end{eqnarray}
and further
\[
\im 1 \otimes \sigma_{\alpha}^{1}\sigma_{\beta}^{2} \in \lie,\quad \alpha,\beta=x,y,z,
\]
by evaluating their commutation relations with $\im 1\otimes \sigma_\alpha^i$ ($\alpha=x,y$).

To prove the complete controllability, we need further to prove
$\im s_{j}^{k} \otimes 1 \in \lie $ and $\im s_j^k \otimes \sigma_{\alpha}^k\in \lie$.
This can be achieved as
\begin{eqnarray}
&\frac{1}{2}\left[\im h_{11,2i} \otimes \sigma_{\alpha}^{1}\sigma_{\beta}^{2}, \im x_{11,2i}
\otimes \sigma_{\alpha}^{1}\sigma_{\beta}^{2}\right]
 = \im y_{11,2i} \otimes 1 \in \lie,\\
&\frac{1}{2}\left[\im h_{11,2i} \otimes \sigma_{\alpha}^{1}\sigma_{\beta}^{2}, \im y_{11,2i}
\otimes \sigma_{\alpha}^{1}\sigma_{\beta}^{2}\right]
 = \im x_{11,2i} \otimes 1 \in \lie,\\
&\frac{1}{2}\left[\im x_{11,2i} \otimes \sigma_{\alpha}^{1}\sigma_{\beta}^{2}, \im y_{11,2i}
\otimes \sigma_{\alpha}^{1}\sigma_{\beta}^{2}\right]
 = \im h_{11,2i} \otimes 1 \in \lie,
\quad i=1,2,
\end{eqnarray}
and
\begin{eqnarray}
& -\frac{1}{2}\left[\im s_j^k\otimes\sigma^1_{\alpha}\sigma^2_{\beta}, \im 1\otimes \sigma_{\gamma}^1\sigma^2_{\beta} \right]
=-\frac{1}{2} \im s_j^k \otimes [\sigma^1_\alpha,\sigma^1_\gamma]=\im s_j^k \otimes \sigma^1_\delta\in\lie,
\label{1sigma}
\end{eqnarray}
where $[\sigma^1_\alpha,\sigma^1_\gamma]=2\im \sigma_\delta$ ($\alpha \neq \gamma$).
$\im s_j^k \otimes \sigma^2_\delta\in\lie$ can be proved similarly.

So far, we have obtained all the elements
$\im s_{j}^{k} \otimes 1 \in \lie$,
$\im s_{j}^{k} \otimes  \sigma_{\alpha}^{1}\sigma_{\beta}^{2} \in \lie$,
$\im 1 \otimes \sigma_{\alpha}^{1}\sigma_{\beta}^{2} \in \lie$
and $\im 1 \otimes \sigma_{\alpha}^{k} \in \lie$. Those elements generate the Lie
algebra $\su(12)$, Therefore, the system is completely controllable
under condition (\ref{eqR:id2det}).

\section{Controllability of a three-level system}

Consider a more complicated system, an 3-level system with $\beta_1=1$,
$\beta_2=\beta_3=2$ and dimension ${\cal N}=5$. 
It is obvious that we need $3$ qubits in the accessor. Then the interaction and accessor Hamiltonians read
\begin{eqnarray}
H_{SA} =
& \sum_{\alpha,\beta,\gamma=x,y,z} \left[ \sum_{n=1}^{2}\sum_{i=1}^{\beta_n}\sum_{j=1}^{\beta_{n+1}}\sum_{k=0,\pm1} g_{\alpha \beta \gamma}^{nij(k)} s_{nij}^{k}\right]
\otimes  \sigma_{\alpha}^{1}\sigma_{\beta}^{2}\sigma_{\gamma}^{3},
\end{eqnarray}
\begin{equation}
H_{A} = \hbar \omega_{1} \sigma_{z}^{1} + \hbar \omega_{2} \sigma_{z}^{2} +\hbar \omega_{3} \sigma_{z}^{3}+
c_{1} \sigma_{x}^{1}\sigma_{x}^{2} +c_{2} \sigma_{x}^{2}\sigma_{x}^{3}.
\end{equation}

From $\im 1 \otimes \sigma_{x}^{k}, \im 1 \otimes \sigma_{y}^{k} \in \lie,
k=1,2,3$, we obtain $\im 1 \otimes \sigma_{z}^{k} \in \lie$. Thus
\begin{eqnarray}
 \im H_{0}^{\prime}=\im H_{0} - \im \sum_{i=1}^{3}\hbar \omega_{i} \sigma_{z}^{i} \in \lie.
\end{eqnarray}
Applying selection operator defined in $(\ref{eqR:idSxy})$, we have
\begin{eqnarray}\label{Sxyz}
& S_{zx}^{3}S_{yx}^{2}S_{xx}^{1}(\im H_0^{\prime}) \nonumber \\
& = \im \left[ \sum_{n=1}^{2}\sum_{i=1}^{\beta_n}\sum_{j=1}^{\beta_{n+1}}\sum_{k=0,\pm1} g_{xyz}^{nij(k)} s_{nij}^{k}\right]
\otimes  \sigma_{x}^{1}\sigma_{x}^{2}\sigma_{x}^{3} \in \lie.
\end{eqnarray}
Similarly, using different combination of $S_{xx}^{k}$, $S_{yx}^{k}$ and $S_{zx}^{k}$, we obtain
\begin{eqnarray}\label{eqR:Ag}
 \im \left[ \sum_{n=1}^{2}\sum_{i=1}^{\beta_n}
 \sum_{j=1}^{\beta_{n+1}}\sum_{k=0,\pm1}
 g_{\alpha \beta \gamma}^{nij(k)} s_{nij}^{k}\right]
\otimes  \sigma_{x}^{1}\sigma_{x}^{2}\sigma_{x}^{3} \in \lie,
\end{eqnarray}
where $\alpha,\beta,\gamma=x,y,z$.

We note that, for the two-level system presented in last section, there is an additional term
$c\sigma_{x}^{1}\sigma_{x}^{2}$ in the element (\ref{the9th-element}),
which cannot be annihilated by the selection operator.
However, for the 3-level system, we need three selection operators to obtain (\ref{eqR:Ag})
and three selection operators always annihilate the term $c\sigma_{x}^{1}\sigma_{x}^{2}$
\begin{equation}
S_{\alpha x}^{1}S_{\beta x}^{2}S_{\gamma x}^{3}(\sigma_{x}^{i}
\sigma_{x}^{i+1})=0,\quad \alpha,\beta,\gamma=x,y,z.
\end{equation}
Therefore we can obtain 27 elements in the form of $(\ref{eqR:Ag})$ and the lower scripts
are $\alpha,\beta,\gamma=x,y,z$.

Total number of elements $\im s_{nij}^{k} \otimes \sigma^1_{x}
\sigma^2_{x}\sigma_{x}^{3}$ in each element of (\ref{eqR:Ag})
is $18$, and we have $3^3=27$ elements of the form (\ref{eqR:Ag}).
Each element can be labeled as $\{ \alpha \beta \gamma\}$ $\alpha,\beta,\gamma=x,y,z$.
To obtain those elements, we need to choose $18$ elements from $(\ref{eqR:Ag})$.
Let $\{\pi\}=\{ \alpha \beta \gamma | \alpha,\beta,\gamma=x,y,z\}$,
the index set of $(\ref{eqR:Ag})$. We can
choose any subset $\{\pi^\prime\}$ of $\{ \pi\}$, containing 18 elements.
Without losing generality, we choose subset $\{ \pi^{\prime}\}$ as
\begin{eqnarray}
\{ \pi^{\prime}\}= & \{ xxx,xxy,xxz,xyx,xyy,xyz,xzx,xzy, xzz, \nonumber \\
& yxx,yxy,yxz,yyx,yyy,yyz,yzx,yzy,yzz  \}.
\end{eqnarray}
If the determinant of coefficient matrix of elements corresponding to $\{\pi^\prime\}$ is
not vanishing, namely
\begin{eqnarray}\label{eqR:id3det}
\det\left( g_{\{ \pi^{\prime}\}}^{nij(k)}\right) \neq 0,
\end{eqnarray}
we have that
\begin{eqnarray}
&\im x_{ni,(n+1)j} \otimes \sigma_{x}^{1}\sigma_{x}^{2}\sigma_{x}^{3} \in \lie, \\
&\im y_{ni,(n+1)j} \otimes \sigma_{x}^{1}\sigma_{x}^{2}\sigma_{x}^{3} \in \lie, \\
&\im h_{ni,(n+1)j} \otimes \sigma_{x}^{1}\sigma_{x}^{2}\sigma_{x}^{3} \in \lie,
\end{eqnarray}
where $n=1,2$, $i=1,2,\cdots,\beta_n$ and
$j=1,2,\cdots,\beta_{n+1}$. Commutation relations of those elements with
$\im 1\otimes \sigma_{\alpha}^{k}$ ($\alpha=x,y,z$) give rise to
\begin{eqnarray}
&\im x_{ni,(n+1)j} \otimes \sigma_{\alpha}^{1}\sigma_{\beta}^{2}\sigma_{\gamma}^{3} \in \lie, \label{51}\\
&\im y_{ni,(n+1)j} \otimes \sigma_{\alpha}^{1}\sigma_{\beta}^{2}\sigma_{\gamma}^{3} \in \lie, \label{52}\\
&\im h_{ni,(n+1)j} \otimes \sigma_{\alpha}^{1}\sigma_{\beta}^{2}\sigma_{\gamma}^{3} \in \lie, \label{53}
\end{eqnarray}
where
$n=1,2$, $i=1,2,\cdots,\beta_n$, $j=1,2,\cdots,\beta_{n+1}$, and $\alpha,\beta,\gamma=x,y,z$.
Without further conditions, we have
\begin{eqnarray}
& \frac{1}{2}\left[\im h_{ni,(n+1)j} \otimes \sigma_{\alpha}^{1}\sigma_{\beta}^{2}\sigma_{\gamma}^{3}, \im x_{ni,(n+1)j} \otimes \sigma_{\alpha}^{1}\sigma_{\beta}^{2}\sigma_{\gamma}^{3}\right] \nonumber \\
& \qquad = \im y_{ni,(n+1)j} \otimes 1 \in \lie,\\
& \frac{1}{2}\left[\im h_{ni,(n+1)j} \otimes \sigma_{\alpha}^{1}\sigma_{\beta}^{2}\sigma_{\gamma}^{3}, \im y_{ni,(n+1)j} \otimes \sigma_{\alpha}^{1}\sigma_{\beta}^{2}\sigma_{\gamma}^{3}\right] \nonumber \\
&\qquad = \im x_{ni,(n+1)j} \otimes 1 \in \lie,\\
&\frac{1}{2}\left[\im x_{ni,(n+1)j} \otimes \sigma_{\alpha}^{1}\sigma_{\beta}^{2}\sigma_{\gamma}^{3}, \im y_{ni,(n+1)j} \otimes \sigma_{\alpha}^{1}\sigma_{\beta}^{2}\sigma_{\gamma}^{3}\right] \nonumber \\
&\qquad = \im h_{ni,(n+1)j} \otimes 1 \in \lie.
\end{eqnarray}

So far, we have obtained
$\im s_{nij}^{k} \otimes \sigma_{\alpha}^{1}\sigma_{\beta}^{2}\sigma_{\gamma}^{3} \in \lie$,
$\im s_{nij}^{k} \otimes 1 \in \lie$ and $\im 1 \otimes \sigma_{\alpha}^{k} \in \lie$. It is easy to
find that $\im s_{nij}^{k} \otimes \sigma_{\alpha}^{l} \in \lie$ as in (\ref{1sigma}).
To prove the complete controllability, we need to further prove
$\im s_{nij}^{k} \otimes \sigma_{\alpha}^{i}\sigma_{\beta}^{j} \in \lie$,
$\im 1 \otimes \sigma_{\alpha}^{i}\sigma_{\beta}^{j} \in \lie \ (i,j=1,2,3)$ and
$\im 1 \otimes \sigma_{\alpha}^{1}\sigma_{\beta}^{2}\sigma_{\gamma}^{3} \in \lie$.
Subtracting elements (\ref{51}-\ref{53}) from $\im H_A^{\prime}$, one finds
\begin{eqnarray}
\im H_A^{I}
&=&  \im H_{0}^{\prime}
 - \im \sum_{\alpha,\beta,\gamma=x,y,z}\left[ \sum_{n=1}^{2}
\sum_{i=1}^{\beta_n}
\sum_{j=1}^{\beta_{n+1}}\sum_{k=0,\pm1} g_{\alpha \beta \gamma}^{nij(k)}
s_{nij}^{k}\right]
 \otimes  \sigma_{\alpha}^{1}\sigma_{\beta}^{2}\sigma_{\gamma}^{3} \nonumber \\
&=& c_1 \im 1 \otimes \sigma_{x}^{1}\sigma_{x}^{2} + c_2 \im 1 \otimes \sigma_{x}^{2}\sigma_{x}^{3}\in \lie.
\end{eqnarray}
Then we have
\begin{eqnarray}
\im 1 \otimes \sigma_{x}^{1}\sigma_{x}^{2} =& -4c_1^{-1}[\im H_A^{I},\im 1
\otimes \sigma_{y}^{1}],\im 1 \otimes \sigma_{y}^{1}] \in \lie , \\
\im 1 \otimes \sigma_{x}^{2}\sigma_{x}^{3} =& c_2^{-1}(\im H_A^{I}, -c_1 \im 1
\otimes \sigma_{x}^{1}\sigma_{x}^{2}) \in \lie .
\end{eqnarray}
From Lemma 2 in \cite{fu-indirect2}, we obtain
$\im 1 \otimes \sigma_{\alpha}^{i}\sigma_{\beta}^{j} \in \lie \ (i,j=1,2,3)$ and
$\im 1 \otimes \sigma_{\alpha}^{1}\sigma_{\beta}^{2}\sigma_{\gamma}^{3} \in \lie$.
Then from $\im s_{nij}^{k} \otimes 1 \in \lie$, we obtain
$\im s_{nij}^{k} \otimes \sigma_{\alpha}^{i}\sigma_{\beta}^{j} \in \lie\ (i,j=1,2,3)$.
We know that we have obtained all needed basis elements, which
generate the Lie algebra $\su\left(2^M {\cal N}\right)=\su(40)$,
and thus the system is completely
controllable if condition $(\ref{eqR:id3det})$ is satisfied.

\section{General case}

With explicit examples in last two sections, we now turn to the indirect control of
arbitrary degenerate quantum system. Assume that $N\geq 3$ and therefore
$M \geq 3$.

At first, from $\im 1 \otimes \sigma_{x}^{k}, \im 1 \otimes \sigma_{y}^{k} \in \lie \ (k=1,2,\cdots,M)$,
we obtain $\im 1 \otimes \sigma_{z}^{k} \in \lie$. Then we have
\begin{eqnarray}
 \im H_{0}^{\prime}=\im H_{0} - \im \sum_{i=1}^{M}\hbar \omega_{i} 1\otimes\sigma_{z}^{i} \in \lie.
\end{eqnarray}
As the coupling terms $\sigma_{x}^{i}\sigma_{x}^{i+1}$ in accessor Hamiltonian are
product of only two Pauli matrices, we have
\begin{equation}
S_{*}^{1}S_{*}^{2}\cdots S_{*}^{M}(\sigma_{x}^{i}\sigma_{x}^{i+1})=0,
\label{selection-kill-ih-prime}
\end{equation}
where $*$ are one of $x, y, z$. So same as previous examples, we apply combination of
selection operators $S_{xx}^{k}$£¬$S_{yx}^{k}$ and $S_{zx}^{k}$ to obtain the following
elements of $\lie$
\begin{eqnarray}\label{eqR:NSxyz}
& S_{\alpha_{M}x}^{M}\cdots S_{\alpha_{2}x}^{2}
S_{\alpha_{1}x}^{1}(\im H_0^{\prime})  \nonumber \\
& \qquad
=
\im \left[ \sum_{n=1}^{N-1}\sum_{i=1}^{\beta_n}\sum_{j=1}^{\beta_{n+1}}
\sum_{k=0,\pm1} g_{\{ \alpha \}}^{nij(k)} s_{nij}^{k}\right]
\otimes  \left( \prod_{p=1}^{M} \sigma_{x}^{p} \right) \in \lie,
\end{eqnarray}
where $\{ \alpha \}= \{ \alpha_{1}, \alpha_{2},\cdots,\alpha_{M}\}$ and $\alpha_{i}=x, y, z$.

There are $\tilde{N}$ elements
$s_{nij}^{k} \otimes \left( \prod_{p=1}^{M} \sigma_{x}^{p} \right)$
in $H_{0}^{\prime}$. To obtain them,
we need $\tilde{N}$ elements of the form
$(\ref{eqR:NSxyz})$. While $(\ref{eqR:NSxyz})$ can supply $3^M$ elements.
Here we have to require the condition (\ref{chain-length-condition}) is
satisfied.
Let
$\{ \pi\}=\{ \alpha_{1} \alpha_{2} \cdots \alpha_{M} | \alpha_{i} = x,y,z\}$,
the index set of element (\ref{eqR:NSxyz}). Without losing generality, we can
choose any subset $\{\pi^\prime\}$ of $\{\pi\}$, containing $\tilde{N}$ elements
in $\{\pi\}$.
Then, if the following condition is satisfied
\begin{eqnarray}\label{eqR:idNdet}
\det \left( g_{\{ \pi^{\prime}\}}^{nij(k)}\right) \neq 0,
\end{eqnarray}
we immediately obtain that
\begin{eqnarray}
&\im x_{ni,(n+1)j} \otimes \sigma_{x}^{1}\sigma_{x}^{2}\cdots\sigma_{x}^{M}  \in \lie, \\
&\im y_{ni,(n+1)j} \otimes \sigma_{x}^{1}\sigma_{x}^{2}\cdots\sigma_{x}^{M} \in \lie, \\
&\im h_{ni,(n+1)j} \otimes \sigma_{x}^{1}\sigma_{x}^{2}\cdots\sigma_{x}^{M} \in \lie,
\end{eqnarray}
where $n=1,2,\cdots,N-1$, $i=1,2,\cdots,\beta_n$ and $j=1,2,\cdots,\beta_{n+1}$.
Commutation relations of those elements with $\im  1 \otimes \sigma_{\alpha_k}^{k}$
($\alpha_k=x,y,z$) give rise to
\begin{eqnarray}
&\im x_{ni,(n+1)j} \otimes \sigma_{\alpha_{1}}^{1}\sigma_{\alpha_{2}}^{2}\cdots\sigma_{\alpha_{M}}^{M}  \in \lie, \\
&\im y_{ni,(n+1)j} \otimes \sigma_{\alpha_{1}}^{1}\sigma_{\alpha_{2}}^{2}\cdots\sigma_{\alpha_{M}}^{M} \in \lie, \\
&\im h_{ni,(n+1)j} \otimes \sigma_{\alpha_{1}}^{1}\sigma_{\alpha_{2}}^{2}\cdots\sigma_{\alpha_{M}}^{M} \in \lie,
\end{eqnarray}
where $n=1,2,\cdots,N-1$, $i=1,2,\cdots,\beta_n$ and
$j=1,2,\cdots,\beta_{n+1}$, $\alpha_{i}=x,y,z$. Without any further condition, we can obtain
\begin{eqnarray}
& \frac{1}{2}\left[\im h_{ni,(n+1)j} \otimes  \sigma_{\alpha_{1}}^{1}
\sigma_{\alpha_{2}}^{2}\cdots\sigma_{\alpha_{M}}^{M} , \im x_{ni,(n+1)j}
\otimes  \sigma_{\alpha_{1}}^{1}\sigma_{\alpha_{2}}^{2}\cdots
\sigma_{\alpha_{M}}^{M} \right] \nonumber \\
& \qquad =
 \im y_{ni,(n+1)j} \otimes 1 \in \lie,\\
& \frac{1}{2}\left[\im h_{ni,(n+1)j} \otimes \sigma_{\alpha_{1}}^{1}
\sigma_{\alpha_{2}}^{2}\cdots\sigma_{\alpha_{M}}^{M}, \im y_{ni,(n+1)j}
\otimes \sigma_{\alpha_{1}}^{1}\sigma_{\alpha_{2}}^{2}\cdots
\sigma_{\alpha_{M}}^{M} \right] \nonumber \\
& \qquad = \im x_{ni,(n+1)j} \otimes 1 \in \lie,\\
& \frac{1}{2}\left[\im x_{ni,(n+1)j} \otimes \sigma_{\alpha_{1}}^{1}
\sigma_{\alpha_{2}}^{2}\cdots\sigma_{\alpha_{M}}^{M}, \im y_{ni,(n+1)j}
\otimes \sigma_{\alpha_{1}}^{1}\sigma_{\alpha_{2}}^{2}\cdots
\sigma_{\alpha_{M}}^{M} \right] \nonumber \\
& \qquad = \im h_{ni,(n+1)j} \otimes 1 \in \lie.
\end{eqnarray}
Further calculation shows that
\begin{eqnarray}
\im H_A^{I} &=& \im H_{0}^{\prime} - \im \sum_{\{ \alpha \}}
\left[ \sum_{n=1}^{N-1}\sum_{i=1}^{\beta_n}\sum_{j=1}^{\beta_{n+1}}
\sum_{k=0,\pm1} g_{\{ \alpha \}}^{nij(k)} s_{nij}^{k}\right]
  \otimes  \left( \prod_{i=1}^{M}\sigma_{\alpha_{i}}^{i} \right) \nonumber \\
&=& \sum_{j=1}^{M-1}c_i \im 1 \otimes \sigma_{x}^{j}\sigma_{x}^{j+1} \in \lie.
\end{eqnarray}
Then we know
\begin{eqnarray}
&&  \im 1 \otimes \sigma_{x}^{1}\sigma_{x}^{2} = -4c_1^{-1}\left[\left[\im H_A^{I}, \im 1 \otimes \sigma_{y}^{1}\right],
    \im 1 \otimes \sigma_{y}^{1}\right] \in \lie , \\
&&  \im 1 \otimes \sigma_{x}^{2}\sigma_{x}^{3} =
 -4c_2^{-1}\left[\left[\im H_A^{I}-c_1 \im 1 \otimes \sigma_{x}^{1}\sigma_{x}^{2}, \im 1
\otimes \sigma_{y}^{2}\right],
 \im 1 \otimes \sigma_{y}^{2}\right] \in \lie ,
\end{eqnarray}
Repeating above two steps, we can obtain
\[
\im 1 \otimes \sigma_{x}^{j}\sigma_{x}^{j+1} \in \lie,\quad j=1,2,\cdots,M-1.
\]
According to Lemma 2 in \cite{fu-indirect2}, we obtain
\begin{eqnarray}
& \im 1 \otimes \left( \prod_{i=1}^{M}\sigma_{\alpha_{i}}^{i} \right) \in \lie ,
\qquad
 \alpha_{i}=0,x,y,z;\ \{ \alpha_{i} \} \neq \{0,0,\cdots,0\},
\end{eqnarray}
where $\sigma_{0}^{i}=1$.
Then from $\im  s_{nij}^{k} \otimes 1 \in \lie$, we obtain
$\im  s_{nij}^{k} \otimes \left( \prod_{i=1}^{M}\sigma_{\alpha_{i}}^{i} \right)
\in \lie , \{ \alpha_{i} \} \neq \{0,0,\cdots,0\}$.

So far, we already have
\begin{eqnarray}
& \im s_{nij}^{k} \otimes \left( \prod_{i=1}^{M}\sigma_{\alpha_{i}}^{i} \right)
  \in \lie \qquad (\{ \alpha_{i} \} \neq \{0,0,\cdots,0\}), \nonumber \\
& \im s_{nij}^{k} \otimes 1 \in \lie, \nonumber \\
& \im 1 \otimes \left( \prod_{i=1}^{M}\sigma_{\alpha_{i}}^{i}
\right) \in \lie \qquad (\{ \alpha_{i} \} \neq \{0,0,\cdots,0\}).
\end{eqnarray}
We have obtained all basis elements of Lie algebra $\su(2^M \cal{N})$,
and the system is thus completely controllable under conditions
(\ref{chain-length-condition}) and (\ref{eqR:idNdet}).

\section{Conclusion}

In this paper, we investigated the complete controllability of
finite dimensional degenerate quantum systems in the indirect
control scheme. The quantum accessor consists of $M$ qubits with
nearest-site coupling. The conditions for length of qubit chain
and the way of coupling between controlled system and accessor
is found. The investigation is illustrated explicitly with two-
and three-level degenerate systems as examples. The degenerate
system is completely controllable if the following conditions
are satisfied
\begin{itemize}
\item The number $M$ of qubits in accessor satisfies condition
      the $3^M \geq \tilde{N}$;

\item Coupling constant matrix (\ref{eqR:idNdet}) between system and accessor
      has non-vanishing determinant.
\end{itemize}

As further works, we would like to generalize the investigation presented
in this paper to the indirect control protocol of
finite quantum systems, and the control protocol of quantum systems in the
presence of environment.

\section*{Acknowledgement}

This work is supported by the National Science Foundation of China under grand
numbers 11075108 and 61374057.

\end{document}